\documentclass{article}

\usepackage{arxiv}

\usepackage[utf8]{inputenc} 
\usepackage[T1]{fontenc}    
\usepackage{hyperref}       
\usepackage{url}            
\usepackage{booktabs}       
\usepackage{amsfonts}       
\usepackage{nicefrac}       
\usepackage{microtype}      
\usepackage{lipsum}
\usepackage{graphicx}

\usepackage{algorithmic}
\usepackage[ruled,vlined]{algorithm2e}
\usepackage{amsthm}
\usepackage{mathrsfs}  
\usepackage{framed}
\usepackage{enumitem}
\usepackage{tcolorbox}
\usepackage{amsmath}

\graphicspath{ {./images/} }

\usepackage{tikz}
\usetikzlibrary{arrows.meta, decorations.pathmorphing}
\usepackage{subcaption}
\tikzset{
  nd/.style     = {circle, draw=gray!70, fill=gray!30, minimum
  size=6mm, inner sep=0pt},
  active/.style = {circle, draw=black,   fill=gray!30, minimum
  size=6mm, inner sep=0pt, line width=1pt},
  new/.style    = {circle, draw=teal!60!black, fill=teal!70!black, text=white,
  minimum size=6mm, inner sep=0pt, font=\bfseries},
  e/.style      = {gray!70},
  se/.style     = {black, line width=1pt, -{Stealth[length=2mm]}},
  bp/.style     = {black, line width=1pt, -{Stealth[length=2mm]}}
}
\newcommand{\mctspanel}[2]{%
  \begin{minipage}[t]{4.6cm}\centering
    \begin{tikzpicture}
      \path[use as bounding box] (-2,-2) rectangle (2,2.3);
      #1
    \end{tikzpicture}\\[4pt]
    {\normalsize \selectfont #2}
\end{minipage}}

\begin{document}

\title{Practical Training-Free MCTS Query Optimization}

\author{
Vladimir Burlakov\\
Lomonosov Moscow State University\\
\texttt{burlakovvs@my.msu.ru}\\
\and
\textbf{Alena Rybakina}\\
Innopolis University\\
\and
\textbf{Sergey Kudashev}\\
Innopolis University\\
\and
\textbf{Konstantin Gilev}\\
Novosibirsk State University\\
\and
\textbf{Alexander Demin}\\
Ershov Institute of Informatics Systems\\
\and
\textbf{Denis Ponomaryov}\\
Ershov Institute of Informatics Systems\\
\and
\textbf{Yuriy Dorn}\\
Lomonosov Moscow State University\\
}

\maketitle

\begin{abstract}
  Production query optimizers typically switch from exhaustive
  dynamic programming to heuristic or randomized search, such as
  greedy ordering or genetic algorithms, once the number of joined
  relations reaches roughly a dozen. This transition trades
  optimality guarantees for
  scalability and can substantially degrade plan quality as the
  join-order search space grows exponentially. Learned alternatives
  such as AlphaJoin and HyperQO can improve search quality, but often
  rely on workload-specific training data, distribution-stable query
  patterns, and deployment mechanisms such as external hint rewriting.

  We present MCTS-Extreme, a training-free Monte Carlo Tree Search (MCTS)
  query optimizer that reuses the host system's existing cost model
  instead of replacing it with learned components.  The system
  combines a subtree-incumbent MCTS selection rule with a
  configurable plan-shape parameter, and runs natively as
  a PostgreSQL planner extension rather than through external hint
  rewriting.

  We evaluate MCTS-Extreme against AlphaJoin, HyperQO, and
  PostgreSQL’s production optimizers, including exhaustive dynamic
  programming and GEQO, on JOB, JOB-Complex, and IMDb-CEB. In the
  high-arity regime where exhaustive enumeration is infeasible,
  MCTS-Extreme produces plans with lower end-to-end response time than
  PostgreSQL's baselines on the high-arity IMDb-family slices, with
  planning time below both baselines at the workload level and
  comparable to GEQO's on high-arity queries.  A $K{=}1$ control
  shows that MCTS outperforms two local-search algorithms restricted
  to the same linear, or zig-zag, plan shape.  In our reproduction
  study, the learned MCTS systems we could run did not improve on the
  default optimizer. We release the PostgreSQL extension source code
  and benchmark artifacts.
\end{abstract}


\section{Introduction}

Cost-based optimizers usually enumerate join orders exactly only for
small queries.  PostgreSQL, for example, switches from dynamic
programming to GEQO once a query reaches the default threshold of
12 joined relations.  That switch keeps planning time bounded, but
it can also leave the optimizer with a plan whose execution time is
far from the best plan available under the same cost model.

In this paper, we revisit MCTS-based approaches to join order
selection.  Methods such as AlphaJoin \cite{zhang2020alphajoin}
and HyperQO \cite{yu2022cost} have shown promising results, but
rely on learned cost models or value networks whose
generalization is fragile and whose deployment depends on
external processes that feed plans back to the planner through
hints.  We propose MCTS-Extreme, a training-free MCTS-based
join optimizer integrated natively into PostgreSQL as a
planner extension.  The algorithm is distinguished by a
UCT-Extreme selection rule that targets the best plan in each
subtree rather than the average plan, and uses the database's own
internal cost model directly.  The main contributions of this
paper are summarized as follows.

\paragraph{Contributions.}
(1) \emph{Deployable MCTS join optimizer.}  We describe and
release MCTS-Extreme, an MCTS-based join optimizer integrated
natively into PostgreSQL as a planner extension.  The extension
uses PostgreSQL's own cost model and is released with complete
source code.
(2) \emph{UCT-Extreme selection rule.}
We give an operational description of the UCT-Extreme selection
rule.  A plan-shape parameter is exposed as a runtime
knob; across the configurations we explored, its $K{=}1$
linear/zig-zag mode delivered the best end-to-end response time
and is the default for every main result in this paper.
(3) \emph{Performance evaluation.}  On the Join Order Benchmark
(JOB), JOB-Complex, and a 200-query IMDb-CEB subset restricted
to $\geq 9$ joined relations, MCTS-Extreme outperforms
PostgreSQL's DPSize and GEQO in the high-arity regime where the
default planner already switches to GEQO.  On JOB queries with
$\geq 12$ joined relations, MCTS-Extreme is $1.47\times$ faster
than GEQO and $1.40\times$ faster than DPSize end-to-end; on the
independent IMDb-CEB $\geq 12$ slice it is $1.17\times$ and
$1.29\times$ faster, respectively.  Planning time is comparable
to GEQO's and roughly $2$--$4\times$ lower than DPSize's on the
high-arity slices.

\section{Background and Related Work}

\subsection{Join Order Optimization Problem}

Join order optimization (JOO) is the task of determining the
optimal sequence of join operations for a set of
relations $R = \{R_1, R_2, \dots, R_n\}$ mentioned in a query
such that the estimated query execution cost is minimized.
Given a query $Q$ involving these relations and a set of join
predicates $P$, the optimizer must select a join tree $T$ from
the space of all semantically equivalent valid join trees
$\mathcal{T}$.

Formally, let $C: \mathcal{T} \to \mathbb{R}^+$ be a cost
function derived from a specific cost model (in practice, a query
  plan cost estimation model implemented in the database that
  accounts for data selectivity, I/O, CPU resource consumption, and
other factors).  The optimization objective is to find:
\begin{equation}
  T_{\text{opt}} = \arg\min_{T \in \mathcal{T}} C(T).
\end{equation}
This problem is non-trivial due to the combinatorial explosion
of the plan search space and the typical inaccuracy of a cost
model (low correlation of cost estimation with actual plan
execution latency).

The search space for the JOO problem grows factorially with the
number of relations.  For a query with $n$ relations, the number
of possible left-deep join trees is $n!$, while the number of
bushy trees is given by the Catalan-related formula
$(2n-2)!/(n-1)!$ \cite{ono1990measuring}.  Consequently,
determining the existence of a join order with required
properties is known to be NP-hard for general join graphs
\cite{ibaraki1984optimal}.  While dynamic programming algorithms
for plan search are typically used in industrial database
engines for small values of $n$ (typically $n \leq 6$--$12$), the
problem is computationally intractable for queries involving a
large number of joins, necessitating the use of heuristics or
randomized algorithms.

Join-order search spaces are also commonly classified by the shape
of the binary join tree \cite{ono1990measuring,steinbrunn1997heuristic}.
Left-deep and right-deep trees restrict one input of each join to be
a base relation on a fixed side.  Zig-zag trees relax the fixed-side
requirement: every join still has at least one base-relation input,
but the accumulated intermediate may appear on either side.  We use
the term \emph{linear trees} for this class, which contains both
left-deep and right-deep trees and remains strictly less general than
bushy trees.

End-to-end (e2e) response time, including both planning and
execution, is the metric exposed to users.  This motivates
heuristics that may return plans with higher estimated execution
cost than exhaustive search would find, but whose planning
complexity scales better with the number of joined relations.

\subsection{Related Work on Join Order Optimization}

\paragraph{Classical Join Order Optimization Algorithms.}  The
foundational framework for solving the JOO problem was
established in the design of the System R optimizer
\cite{selinger1979access}.  This approach utilized dynamic
programming to construct an optimal plan bottom-up, pruning
sub-optimal paths based on cost estimates.  To keep optimization
time manageable, traditional implementations (like early
versions of DB2 and PostgreSQL) restricted the search space to
left-deep trees, where the right child of a join is always a
base table.  While this reduces complexity, it misses
potentially optimal plans.  This dynamic programming approach
remains the standard exact strategy for queries with a small
number of joins; PostgreSQL's variant is DPSize.

Although effective for smaller queries, dynamic programming is
not scalable to arbitrary queries.  To handle complex queries,
classical optimizers switch to heuristic or randomized
algorithms.  Algorithms such as Greedy Operator Ordering (GOO)
\cite{fegaras1998new} select a pair of relations with the lowest
immediate join cost (or highest selectivity) to join next.
While fast, they often get trapped in local minima because they
make locally optimal choices that may prevent globally optimal
structures (like specific sort orders) later in the plan.
Systems such as PostgreSQL employ GEQO
\cite{steinbrunn1997heuristic} once the number of tables reaches
a threshold (default $12$).

Classical algorithms do not optimize execution time directly;
rather, they optimize the estimated plan cost, a mathematical proxy
for runtime.  The accuracy of
the cost model relies entirely on estimating the number of rows
(cardinality) output by each operator.  Traditional estimators
use histograms and Most Common Value (MCV) lists.  To simplify
calculations, classical optimizers typically assume that the
data distribution is uniform and that predicates on different
columns are statistically independent.

Despite their maturity, classical database query optimizers have
structural limitations that motivate interest in learning-based
methods.  First, traditional optimizers keep no memory of execution
feedback and cannot improve from their mistakes.
Second, the independence assumptions mentioned above frequently
fail in real-world data where correlations are high.  This
leads to large underestimation of intermediate result sizes; a
multi-join estimate can easily be off by orders of magnitude
(for example, estimating one row when the reality is one
million), causing the optimizer to select a Nested Loop Join
instead of a Hash Join, leading to severe performance
regressions.  Finally, cost models are rigid formulas that may
not reflect the current hardware state (I/O and CPU load) and
require manual tuning by database administrators to align with
reality.

\paragraph{AI for Join Order Optimization.}  Learned query
optimization has been studied both as a way to replace error-prone
optimizer components, such as cardinality or cost estimation
\cite{hilprecht13deepdb, kipf2018learned, yang2020neurocard,
marcus12plan}, and as a way to replace or steer join-order
enumeration itself \cite{lehmann2024your, qiao2025learning,
tsesmelis2022database, anneser2023autosteer}.  Examples include
deep reinforcement-learning systems such as ReJoin
\cite{marcus2018deep} and Neo \cite{marcus12neo}, and hybrid
systems such as Bao \cite{marcus2021bao} that steer existing
optimizers rather than replacing them entirely.  More recent work
also applies actor-critic, retrieval, and LLM-based methods to query
optimization \cite{yan2023join, chen2022efficient,
xiong2024autoquo, qiao2025acjoin, liu2025sefrqo,
zhou2025survey, yao2025query, tan2025can}.

MCTS-based join optimizers form the closest line to our work.
These methods use lookahead rollouts to estimate the downstream value
of a join decision, rather than committing greedily to the cheapest
immediate join.  AlphaJoin \cite{zhang2020alphajoin}, RTOS
\cite{yu2020reinforcement}, and HyperQO \cite{yu2022cost} combine
MCTS-style search with learned models; LOOPLINE
\cite{xiong2025loopline} is a parallel MCTS variant with
knowledge-constraint pruning.

One ingredient from the broader algorithms literature directly
informs our design.  Extreme-value variants
of UCT \cite{coulom2007efficient, imagawa2016enhancements,
huangimproving} replace the average-reward statistic of
classical UCT with a maximum-of-rewards statistic, which is more
appropriate for optimization problems where the objective is the
best leaf reward rather than the average.  MCTS-Extreme builds
on this: its selection rule is a UCT-Extreme variant described in
Section~\ref{sec:algorithm}.

\section{Motivation: Deployability Limits of Learned MCTS Optimizers}
\label{sec:motivation}

Our design choices are driven by deployability and
reproducibility.  To motivate them, we describe why two recent
learned MCTS optimizers, AlphaJoin and HyperQO, do not deploy
straightforwardly.  The observations below indicate that
the obstacles are structural rather than incidental to either
implementation.

\paragraph{AlphaJoin.}  The proposed approach is well-motivated,
but the released artifact targets an older PostgreSQL version
and we encountered several practical obstacles when running it on
a modern PostgreSQL build.  Reproducing the published
results required regenerating training data through a Java
pipeline that is sensitive to modern PostgreSQL configuration,
preprocessing the emitted bracketed-hint format so it parses,
and re-training because trained model checkpoints are not
included in the artifact.  Even after these adjustments, the
learned value network does not transfer to out-of-template
queries: on JOB-Complex, whose schema matches the training
schema but whose predicate structure differs from JOB, the
encoded feature dimensionality is inconsistent with the
re-trained model.  We were unable to reproduce the originally
reported performance on either workload in our setup.

\paragraph{HyperQO.}  The released artifact assumes CUDA and
relies on JOB's specific alias naming convention; running it on
JOB-Complex required manual alias normalization on our side.
HyperQO is structured as a steering layer that recommends a
two-relation leading-edge hint and otherwise defers to the
existing optimizer, so its impact is bounded by how often the
two-relation prefix is the deciding factor.  In our evaluation,
$86.75\%$ of JOB queries fall back to the default PostgreSQL
plan because HyperQO did not produce a usable hint, rising to
$100\%$ on JOB-Complex; the mean per-query execution-time
improvement over the default optimizer is only $\sim 1.3\%$, well
below the reduction originally reported, and its end-to-end time
is in fact higher once HyperQO's per-query inference overhead is
included (we do not charge its training cost).

\paragraph{LOOPLINE and RTOS.}  Two earlier MCTS-based optimizers
also belong in this discussion.  RTOS \cite{yu2020reinforcement}
combines tree-LSTM value estimation with MCTS-style rollouts;
LOOPLINE \cite{xiong2025loopline} is a parallel MCTS variant with
knowledge-constraint pruning.  Neither is included as a
quantitative baseline in this paper.  LOOPLINE has no public
artifact available to test from, so we cannot run it on our
engine.  We likewise could not locate a public RTOS artifact or
model checkpoint suitable for reproduction.  We therefore exclude
both systems from the quantitative comparison because there is no
implementation path we can test directly on our PostgreSQL build,
rather than because their search algorithms failed in our setup.

\paragraph{Common Pattern.}  Both reproduced systems are external
Python processes whose effect on the planner is mediated by hint
rewriting.  That mediation layer is what fails first: dependency
drift, hint representation differences, and unfaithful enforcement
of hinted plans collectively dominate the integration burden.
This motivates the design choice that drives the present paper:
place the search \emph{inside} the planner so that the search's
output is the plan the executor follows, with no rewriting and no
fallback ambiguity.  A detailed reproducibility narrative is
included with the artifact materials; this section uses it as
motivation only.

\section{MCTS-Extreme: Algorithm and PostgreSQL Integration}
\label{sec:algorithm}

MCTS represents search as repeated Selection, Expansion, Rollout,
and Backpropagation steps (Fig.~\ref{fig:mcts_scheme}).
In join-order search, a root-to-leaf path is a join order, a
terminal node is a feasible plan, and the propagated value is the
plan cost.  MCTS-Extreme instantiates the four steps with a
cost-based selection rule, shape-constrained actions, randomized
rollouts, and PostgreSQL-native cost evaluation.

\begin{figure}[t]
  \centering
  \resizebox{\linewidth}{!}{%
    \begin{minipage}{19.5cm}\centering
      \mctspanel{
        \node[nd] (r) at (0,2)    {};
        \node[nd] (a) at (-1,1)   {};
        \node[active] (b) at (1,1)  {};
        \node[nd] (c) at (-1.5,0) {};
        \node[nd] (d) at (-0.5,0) {};
        \node[active] (e) at (1,0)  {};
        \node[nd] (f) at (1.7,0)  {};
        \draw[e] (r)--(a) (a)--(c) (a)--(d) (b)--(f);
        \draw[se] (r)--(b);
        \draw[se] (b)--(e);
      }{\textbf{1. Selection}\\Select child by score}\hfill
      \mctspanel{
        \node[nd] (r) at (0,2)    {};
        \node[nd] (a) at (-1,1)   {};
        \node[nd] (b) at (1,1)    {};
        \node[nd] (c) at (-1.5,0) {};
        \node[nd] (d) at (-0.5,0) {};
        \node[nd] (ee) at (0.5,0) {};
        \node[nd] (f) at (1.5,0)  {};
        \node[new] (g) at (1.5,-1){+};
        \draw[e] (r)--(a) (r)--(b) (a)--(c) (a)--(d) (b)--(ee)
        (b)--(f) (f)--(g);
      }{\textbf{2. Expansion}\\Add new child node}\hfill
      \mctspanel{
        \node[nd]   (r) at (0,2)    {};
        \node[nd]   (a) at (-1,1)   {};
        \node[active] (b) at (1,1)  {};
        \node[nd]   (c) at (-1.5,0) {};
        \node[nd]   (d) at (-0.5,0) {};
        \node[active] (e) at (1,0)  {};
        \node[nd]   (f) at (1.7,0)  {};
        \node[active] (g) at (1,-1) {};
        \node[nd, minimum width=10mm] (end) at (-1,-1.4) {\scriptsize End};
        \draw[e] (r)--(a) (a)--(c) (a)--(d) (b)--(f);
        \draw[se] (r)--(b) (b)--(e) (e)--(g);
        \draw[gray!70, decorate, decoration={snake, amplitude=.5mm,
          segment length=2mm},
        -{Stealth[length=2mm]}] (g) to[bend right=20] (end);
      }{\textbf{3. Rollout}\\Complete random plan}\hfill
      \mctspanel{
        \node[nd]   (r) at (0,2)    {};
        \node[nd]   (a) at (-1,1)   {};
        \node[active] (b) at (1,1)  {};
        \node[nd]   (c) at (-1.5,0) {};
        \node[nd]   (d) at (-0.5,0) {};
        \node[active] (e) at (1,0)  {};
        \node[nd]   (f) at (1.7,0)  {};
        \node[active] (g) at (1,-1) {};
        \draw[e] (r)--(a) (a)--(c) (a)--(d) (b)--(f);
        \draw[bp] (g)--(e) (e)--(b) (b)--(r);
      }{\textbf{4. Backpropagation}\\Update path statistics}
    \end{minipage}%
  }
  \caption{Monte Carlo Tree Search loop.}
  \label{fig:mcts_scheme}
\end{figure}

\emph{Notation summary.}  Throughout this section, $n$ is the
relation count of a single query, $B$ is the iteration budget, and
$C^\star$ denotes the incumbent cost (the best plan cost found so
far).  For a search
tree node $j$, $n_j$ is its visit count and $N$ is the visit count
of its parent.  $c$ and $\gamma$ are the UCT exploration constant
and exploration exponent.

\subsection{Architecture and Integration}

MCTS-Extreme installs as a PostgreSQL extension that intercepts the
planner's join-order search.  For a query with at least $n_{\min}$
base relations, the extension runs the MCTS loop described in the
rest of this section and returns the chosen join tree; for smaller
queries it delegates to the default join search.  All cost
evaluations during the search use PostgreSQL's native cost model
without modification, so the extension shares the planner's
cardinality estimates, selectivity calculations, and join-method
choices.  The integration adds no inter-process communication, no
hint rewriting, and no learned components, and is enabled at the
session or cluster level by loading the extension.

\subsection{UCT-Extreme Selection Rule}
\label{sec:reward}

Classical UCT \cite{kocsis2006bandit} balances exploration and
exploitation by selecting the child $j$ that maximizes the index
\begin{equation}
  \mathrm{UCT}_j = \bar{X}_j + c \sqrt{\frac{2 \ln N}{n_j}},
  \label{eq:uct-classical}
\end{equation}
where $\bar{X}_j$ is the empirical mean reward of child $j$, $N$
is the total number of visits to the parent node, $n_j$ is the
number of visits to child $j$, and $c$ is the exploration
constant.  This index is appropriate when the objective is the
average reward of an arm, as in stochastic multi-armed bandits.

For combinatorial optimization problems such as join ordering,
the objective is the best plan in each subtree, not the average
plan.  We therefore use a UCT-Extreme variant
\cite{huangimproving} in which the per-arm statistic is the best
rollout reward seen so far, $\hat{Q}_j = \max_{1 \leq t \leq
n_j} X_{j,t}$, and the index is
\begin{equation}
  \mathrm{UCT}^{\mathrm{ext}}_j = \hat{Q}_j + 2c \left(
  \frac{\ln N}{n_j} \right)^{\gamma},
  \label{eq:uct-extreme-huang}
\end{equation}
where $\gamma$ controls the exploration rate for under-visited
arms.  Equation~\eqref{eq:uct-extreme-huang} replaces the
running mean $\bar{X}_j$ with the running maximum $\hat{Q}_j$ and
raises the exploration bonus to a power.

In MCTS-Extreme we use Equation~\eqref{eq:uct-extreme-huang} but
substitute a different per-arm statistic.  Each node stores the
subtree incumbent $\widehat{C}_j = \min_{1 \leq t \leq n_j}
C_{j,t}$, the minimum cost across all rollouts in the subtree
rooted at child $j$, and the search tracks the global incumbent
$C^\star = \min_j \widehat{C}_j$.  A \emph{reward map} $\phi$
converts the subtree incumbent into a scalar $\hat{r}_j =
\phi(\widehat{C}_j)$ that takes the role of $\hat{Q}_j$ in the
UCT-Extreme index.  Of the reward maps we implement (below), the one used
for every result reported in this paper is the negative log cost,
\begin{equation}
  \hat{r}_j = -\log \widehat{C}_j,
  \label{eq:reward-neg-log}
\end{equation}
which is monotone decreasing in plan cost and compares
multiplicative cost differences.  The selection score is then
\begin{equation}
  \text{Score}_j = \hat{r}_j + \left( c \cdot \frac{\ln N}{n_j}
  \right)^{\gamma},
  \label{eq:uct-extreme}
\end{equation}
which adapts the UCT-Extreme index of
Equation~\eqref{eq:uct-extreme-huang}: the maximum-reward
statistic $\hat{Q}_j$ is replaced by the reward map $\hat{r}_j$,
and the exploration term keeps the polynomial form $(\cdot)^\gamma$
with the constant $c$ folded inside.  Unvisited children receive
a sentinel score so that they are selected before any visited
child.

The extension implements three reward maps.  Letting
$C_{\min}, C_{\max}$ denote the empirical cost envelope observed
during the current search, the maps are:
\begin{itemize}\itemsep1pt
  \item \textsc{neg\_log} (default, used throughout this paper):
    $\hat{r}_j = -\log \widehat{C}_j$;
  \item \textsc{neg\_cost}:
    $\hat{r}_j = -\widehat{C}_j$;
  \item \textsc{norm\_neg\_log}:
    $\hat{r}_j = \bigl(\log C_{\max} - \log \widehat{C}_j\bigr) /
    \bigl(\log C_{\max} - \log C_{\min}\bigr) \in [0,1]$.
\end{itemize}
All three maps are monotone in plan quality and differ only in
how they normalize: by the empirical log-cost envelope
(\textsc{norm\_neg\_log}) or by no envelope at all
(\textsc{neg\_log}, \textsc{neg\_cost}).
Section~\ref{sec:ablation} compares the three maps.  All reported
benchmarks use \textsc{neg\_log}.

\subsection{Plan-Shape Mode}
\label{sec:planshape}

MCTS-Extreme uses a plan-shape parameter $K \geq 0$.  $K{=}0$
permits unrestricted bushy joins within the MCTS action set.  For
$K{\geq}1$, each rollout completes the expanded partial plan within
what we call a bounded-bushy $K$-linear-component search space, or
$K$-component space for short.  The remaining relations are
organized into at most $K$ connected linear components, a legal
linear chain is constructed inside each component, and the resulting
component outputs are joined by a binary merge tree.  Thus $K$
controls the amount of bushiness.

For $K{=}1$, the search constructs a linear join tree.  Because the
shape predicate allows the accumulated intermediate to appear on
either input side of the next join, this is the standard zig-zag
class rather than a strict left-deep or right-deep restriction.  If
an implementation fixed the accumulated intermediate to one side,
the same $K{=}1$ idea would reduce to left-deep or right-deep
enumeration.  For $K{>}1$, the search may build multiple linear
components before merging them, interpolating between linear and
fully bushy enumeration.  Conceptually, if singleton components are
allowed, $K{=}n$ can represent an arbitrary bushy tree by treating
each base relation as its own component; in the extension,
$K{=}0$ is the direct unrestricted-bushy mode.  We use $K{=}1$ as
the default; Section~\ref{sec:ablation} evaluates alternative
values.

A single shape predicate decides whether a candidate join pair is
admissible under the current shape state, and the same predicate is
used during both expansion and rollout so the shape constraint is
enforced uniformly.

\subsection{Rollout Strategy}

Each iteration of the search ends with one or more random
rollouts from the expanded leaf to a complete join order.
Rollouts draw from a single random stream seeded from the base
seed.  They enforce the current shape mode: in the $K{=}1$ mode,
the rollout continues the current linear/zig-zag tree, while for
$K{>}1$ it builds the allowed $K$-component tree and then
merges the component outputs, with relaxed-shape fallbacks to
guarantee termination on all join graphs.

\subsection{Other Parameters}

Two further parameters bound planning time.  A top-$k$ filter
limits the per-state action set to the $k$ cheapest candidate
joins.  A tree-depth bound determines when a leaf is rolled out
rather than expanded.  The search also caches join costs so that
the same partial plan is not re-evaluated.

\subsection{Algorithm Summary}

Algorithm~\ref{alg:mcts-extreme} summarizes the search procedure.
The search
builds one tree $\mathcal{V}$ backed by a cost cache $\mathcal{H}$,
draws its random stream from the base seed $\sigma_0$, and runs up
to $B$ MCTS iterations.  Within an iteration,
the four standard MCTS steps execute as instantiated above:
Selection follows the UCT-Extreme score
(Equation~\eqref{eq:uct-extreme}), Expansion picks a
shape-admissible action from the top-$k$ candidates under mode
$K$, Rollout completes the partial plan to a terminal node, and
Backpropagation updates the subtree incumbent $\widehat{C}_w$
along the path.  The global incumbent $C^\star$ and the best
join tree $T^\star$ track the best plan found across all $B$
iterations.

\begin{algorithm}[t]
  \caption{MCTS-Extreme: Join Order Search.}
  \label{alg:mcts-extreme}
  \DontPrintSemicolon
  \SetKwInOut{Input}{Input}
  \SetKwInOut{Output}{Output}
  \Input{Query with $n$ base relations; iteration budget $B$;
    constants $c, \gamma$; plan-shape mode $K$; rollout mode; base seed
  $\sigma_0$.}
  \Output{Best join tree $T^\star$ and cost $C^\star$.}
  $C^\star \gets +\infty$;\quad $T^\star \gets \emptyset$\;
  $\mathcal{V} \gets \{v_0\}$;\quad
  $\mathcal{H} \gets \emptyset$\textit{ // search tree, cost cache}\;
  \For{$t = 1, 2, \dots, B$}{
    \textit{// Selection.}\;
    $v \gets v_0$;\quad
    \While{$v$ fully expanded $\wedge$ $v$ non-terminal}{
      $v \gets \arg\max_{u \in \mathrm{children}(v)}
      \mathrm{Score}(u)$\textit{ // Eq.~\eqref{eq:uct-extreme}}\;
    }
    \textit{// Expansion.}\;
    \If{$v$ non-terminal}{
      pick untried shape-admissible action $a$ from top-$k$ candidates\;
      $u \gets$ new child of $v$ via $a$;\quad attach to
      $\mathcal{V}$;\quad $v \gets u$\;
    }
    \textit{// Rollout.}\;
    $v_\bot \gets$ random terminal completion from $v$ under mode
    $K$, seed $\sigma_0$\;
    $C \gets \mathrm{Cost}(v_\bot)$\textit{ // PostgreSQL cost
    model, cached in $\mathcal{H}$}\;
    \textit{// Backprop.}\;
    \For{$w$ on path $v_\bot \to v_0$}{
      $\widehat{C}_w \gets \min(\widehat{C}_w,\, C)$;\quad
      $n_w \gets n_w + 1$\;
    }
    \If{$C < C^\star$}{
      $C^\star \gets C$;\quad $T^\star \gets$ tree of $v_\bot$\;
    }
  }
  \Return $T^\star$
\end{algorithm}

\section{Experimental Evaluation}
\label{sec:experiments}

We evaluate MCTS-Extreme on the Join Order Benchmark (JOB)
\cite{leis2015good}, JOB-Complex \cite{wehrstein2150job}, and
IMDb-CEB \cite{negi2023robust} against PostgreSQL's two production
join-search engines: the
dynamic-programming planner DPSize, and the genetic optimizer
GEQO.  The primary endpoint is aggregate end-to-end response
time (planning plus execution) relative to GEQO on each workload:
the GEQO workload total divided by the MCTS-Extreme workload
total.  DPSize comparisons, subset analyses, geometric means, and
planning-time fractions are reported as secondary measurements.

Both production baselines run in their default regimes:
PostgreSQL switches from DPSize to GEQO once the number of
joined relations reaches the default \texttt{geqo\_threshold} of
12, so for queries with
$\geq 12$ joined relations the production-relevant baseline is
GEQO; the DPSize numbers we report in that regime are included
for completeness and not as a deployment comparison.
Section~\ref{sec:learned-comparison} reports outcomes for AlphaJoin
and HyperQO under the reproduction setup of
Section~\ref{sec:motivation}.  We do not charge their training cost,
Section~\ref{sec:discussion} discusses it separately.

\subsection{Setup}

All experiments ran on a compute node with an AMD EPYC~9654
(Zen~4) processor providing 96 vCPUs at a $2.4$\,GHz base
frequency and $192$\,GB of RAM, under Ubuntu~24.04 LTS.  We use
PostgreSQL built from the \texttt{master} branch
(\texttt{19devel}, February 2026) with MCTS-Extreme installed
through the planner hook.  We use the IMDb dataset and
schema published with JOB, the 113 JOB queries, the 30
JOB-Complex queries, and a
reproducible 200-query subset of IMDb-CEB \cite{negi2023robust}
sampled with seed $42$ and restricted to queries with $\geq 9$
joined relations (Section~\ref{sec:ceb}).  Cold OS page cache is
enforced between queries; the per-query
\texttt{statement\_timeout} is 5\,min.
Every cell (query $\times$ optimizer) is repeated with 3 random
seeds, and for each cell we report the mean over seeds of the
chosen plan cost, planning time, execution time, and end-to-end
response time (planning${+}$execution).

\smallskip\noindent\emph{Why IMDb-based workloads.}\;
We evaluate on the high-arity regime: queries that join $12$ or
more relations.  This is where PostgreSQL stops using exact
dynamic programming and switches to GEQO, at the default
\texttt{geqo\_threshold}${=}12$, so it is also where a better
search is most likely to help.  Public real-data benchmarks that
exercise this regime are scarce.  TPC-H has only $8$ base tables,
so no query joins $12$ relations and GEQO never runs; below the
threshold PostgreSQL already finds the optimal plan with exact DP.
TPC-DS is larger, but its generated data is weakly correlated, so
cardinality estimates stay accurate and join ordering stays easy
even on bigger queries.  The hard case we target is estimation
error that grows across many joins on real, correlated data, which
synthetic generators are not designed to stress.  JOB,
JOB-Complex, and IMDb-CEB constitute the public real-data
benchmark family that reaches this regime.  All use the IMDb
schema, we call them the IMDb family and treat this as a
benchmark-availability limitation rather than evidence of
cross-domain generality.

Unless stated otherwise, MCTS-Extreme uses $B=480$ iterations,
$c=1.41$, $\gamma=0.5$, $K=1$, top-$k$ filtering with $k=10$,
depth bound $d=10$, and one random rollout per leaf.
Section~\ref{sec:ablation} evaluates plan shape, selection rule,
reward map, and search budget on IMDb-CEB.  The reported planning
times are for this configuration; larger budgets or wider action
sets increase planning time.

\smallskip\noindent\emph{Uncertainty quantification.}\;
Each cell runs with three independent seeds.  Each
optimizer invocation uses one seed, so we report the mean over the
three seeds and quantify uncertainty across queries.  For each
aggregate we report a $95\%$ confidence interval from a paired
bootstrap over queries ($10^4$ resamples, recomputing the
ratio-of-sums speedup) and a two-sided Wilcoxon signed-rank test
on the paired per-query e2e times (Table~\ref{tab:significance}).
The seed-to-seed spread is small: on every seed the JOB $\geq 12$
speedup stays above $1.25\times$ against both baselines, so the
main JOB high-arity result does not depend on a particular seed.

\subsection{Aggregate Results}

Tables~\ref{tab:workload_aggregates}--\ref{tab:significance}
separate the production comparison from diagnostic baselines.
The production question is GEQO versus MCTS-Extreme above
PostgreSQL's GEQO threshold; DPSize on those slices is a useful
reference point, but not the mode PostgreSQL would normally choose.
On this comparison, the evidence is strongest on JOB and IMDb-CEB:
both high-arity slices improve over GEQO with confidence intervals
that exclude parity.  JOB-Complex is less decisive.  It remains a
useful stress test, but its high-arity GEQO comparison is too small
and too noisy to support a separate significance claim.

The tables also show why estimated plan cost cannot be treated as
the primary endpoint.  Several MCTS-Extreme plans are rated no
better, or slightly worse, by PostgreSQL's own cost model while
still reducing execution time.  The value of the search is therefore
not that it learns a new cost model; it is that randomized search
finds native-PostgreSQL plans outside the basin reached by the
default high-arity search.

\begin{table}[!t]
  \centering
  \small
  \setlength{\tabcolsep}{0pt}
  \caption{Full-workload aggregate results. MCTS-E denotes
  MCTS-Extreme; positive gap values mean MCTS-Extreme is faster or
  lower-cost than the baseline.}
  \label{tab:workload_aggregates}
  \begin{tabular*}{\columnwidth}{@{\extracolsep{\fill}}lrrrrr@{}}
    \toprule
    & \multicolumn{3}{c}{Optimizer} & \multicolumn{2}{c}{MCTS-E gap (\%)} \\
    \cmidrule(lr){2-4} \cmidrule(lr){5-6}
    & \multicolumn{1}{c}{DPSize}
    & \multicolumn{1}{c}{GEQO}
    & \multicolumn{1}{c}{MCTS-E}
    & \multicolumn{1}{c}{vs.\ DPSize}
    & \multicolumn{1}{c}{vs.\ GEQO} \\
    \midrule
    \multicolumn{6}{l}{\emph{JOB} (113 queries)} \\
    Total e2e (s)            & 4075.4 & 4029.3 & 3580.2 & $+14\%$ & $+13\%$ \\
    Total execution (s)      & 4051.7 & 4013.4 & 3565.8 & $+14\%$ & $+13\%$ \\
    Total plan cost ($10^3$) & 4570   & 4459   & 4616   & $-1\%$ & $-3\%$ \\
    Mean planning (ms)       &  209.2 &  140.6 &  127.4 & $+64\%$ & $+10\%$ \\
    \midrule
    \multicolumn{6}{l}{\emph{JOB-Complex} (29 queries)} \\
    Total e2e (s)            & 2096.2 & 1965.5 & 1806.4 & $+16\%$ & $+9\%$ \\
    Total execution (s)      & 2087.0 & 1960.6 & 1801.8 & $+16\%$ & $+9\%$ \\
    Total plan cost ($10^3$) & 1025   & 1017   & 1010   & $+2\%$ & $+1\%$ \\
    Mean planning (ms)       &  320.3 &  168.5 &  157.6 & $+103\%$ & $+7\%$ \\
    \midrule
    \multicolumn{6}{l}{\emph{IMDb-CEB rel9 subset} (200 queries)} \\
    Total e2e (s)            & 8387.7 & 7722.1 & 7271.2 & $+15\%$ & $+6\%$ \\
    Total execution (s)      & 8340.7 & 7685.0 & 7235.0 & $+15\%$ & $+6\%$ \\
    Total plan cost ($10^3$) &35897   &32151   &33262   & $+8\%$ & $-3\%$ \\
    Mean planning (ms)       &  235.0 &  185.5 &  180.8 & $+30\%$ & $+3\%$ \\
    \bottomrule
  \end{tabular*}
\end{table}

\begin{table}[!t]
  \centering
  \small
  \setlength{\tabcolsep}{0pt}
  \caption{High-arity aggregate results. MCTS-E denotes
  MCTS-Extreme; rels denotes joined relations; positive gap values
  mean MCTS-Extreme is faster or lower-cost than the baseline.}
  \label{tab:high_arity_aggregates}
  \begin{tabular*}{\columnwidth}{@{\extracolsep{\fill}}lrrrrr@{}}
    \toprule
    & \multicolumn{3}{c}{Optimizer} & \multicolumn{2}{c}{MCTS-E gap (\%)} \\
    \cmidrule(lr){2-4} \cmidrule(lr){5-6}
    & \multicolumn{1}{c}{DPSize}
    & \multicolumn{1}{c}{GEQO}
    & \multicolumn{1}{c}{MCTS-E}
    & \multicolumn{1}{c}{vs.\ DPSize}
    & \multicolumn{1}{c}{vs.\ GEQO} \\
    \midrule
    \multicolumn{6}{l}{\emph{JOB} $\cap \{\text{rels} \geq 12\}$ (20
    queries)} \\
    Total e2e (s)            &  443.5 &  465.4 &  316.6 & $+40\%$ & $+47\%$ \\
    Total execution (s)      &  430.6 &  461.9 &  313.0 & $+38\%$ & $+48\%$ \\
    Total plan cost ($10^3$) &  115   &  122   &  124   & $-8\%$ & $-2\%$ \\
    Mean planning (ms)       &  645.2 &  172.9 &  177.0 & $+264\%$ & $-2\%$ \\
    \midrule
    \multicolumn{6}{l}{\emph{JOB-Complex} $\cap \{\text{rels} \geq
    12\}$ (12 queries)} \\
    Total e2e (s)            &  930.3 &  724.6 &  632.6 & $+47\%$ & $+15\%$ \\
    Total execution (s)      &  923.2 &  722.2 &  630.1 & $+47\%$ & $+15\%$ \\
    Total plan cost ($10^3$) &  563   &  581   &  573   & $-2\%$ & $+1\%$ \\
    Mean planning (ms)       &  591.3 &  198.8 &  204.9 & $+189\%$ & $-3\%$ \\
    \midrule
    \multicolumn{6}{l}{\emph{IMDb-CEB} $\cap \{\text{rels} \geq 12\}$
    (57 queries)} \\
    Total e2e (s)            & 3527.5 & 3193.0 & 2739.1 & $+29\%$ & $+17\%$ \\
    Total execution (s)      & 3501.5 & 3180.6 & 2726.1 & $+28\%$ & $+17\%$ \\
    Total plan cost ($10^3$) & 9682   & 7369   & 7426   & $+30\%$ & $-1\%$ \\
    Mean planning (ms)       &  455.5 &  217.0 &  229.2 & $+99\%$ & $-5\%$ \\
    \bottomrule
  \end{tabular*}
\end{table}

\begin{table}[!t]
  \centering
  \small
  \setlength{\tabcolsep}{0pt}
  \caption{High-arity speedup significance. Bold confidence
  intervals exclude parity.}
  \label{tab:significance}
  \begin{tabular*}{\columnwidth}{@{\extracolsep{\fill}}llrcr@{}}
    \toprule
    Slice & Base
    & \multicolumn{1}{c}{Speedup}
    & \multicolumn{1}{c}{$95\%$ CI}
    & \multicolumn{1}{c}{$p$} \\
    \midrule
    JOB $\geq 12$ (20)        & GEQO   & $1.47\times$ &
    \textbf{[1.17, 1.82]} & $0.04$ \\
    & DPSize & $1.40\times$ & \textbf{[1.13, 1.85]} & $0.02$ \\
    \midrule
    JOB-Complex $\geq 12$ (12) & GEQO   & $1.15\times$ & [0.86, 1.55]
    & $0.23$ \\
    & DPSize & $1.47\times$ & \textbf{[1.19, 1.77]} & $0.03$ \\
    \midrule
    JOB${\cup}$Complex $\geq 12$ (32) & GEQO   & $1.25\times$ &
    \textbf{[1.02, 1.54]} & $0.01$ \\
    & DPSize & $1.45\times$ & \textbf{[1.23, 1.68]} & $0.001$ \\
    \midrule
    IMDb-CEB $\geq 12$ (57)   & GEQO   & $1.17\times$ &
    \textbf{[1.08, 1.26]} & $0.001$ \\
    & DPSize & $1.29\times$ & \textbf{[1.13, 1.48]} & $0.004$ \\
    \midrule
    IMDb-CEB $\geq 14$ (28)   & GEQO   & $1.32\times$ &
    \textbf{[1.19, 1.48]} & ${<}10^{-4}$ \\
    & DPSize & $1.42\times$ & \textbf{[1.25, 1.65]} & ${<}10^{-5}$ \\
    \bottomrule
  \end{tabular*}
\end{table}

Table~\ref{tab:winners} gives a per-query view of the same story.
MCTS-Extreme is most often among the best optimizers, but the tie
band matters: many wins fall inside the $\pm 5\%$ tie band rather
than representing large per-query improvements.  The aggregate gains
come from avoiding bad high-arity plans often enough, not from
dominating every query.

\begin{table}[!t]
  \centering
  \small
  \setlength{\tabcolsep}{0pt}
  \caption{Per-query co-winner counts. An optimizer is a co-winner
  on a query if its mean e2e time is within the $\pm 5\%$ tie band
  of the best optimizer's.}
  \label{tab:winners}
  \begin{tabular*}{\columnwidth}{@{\extracolsep{\fill}}lrrrr@{}}
    \toprule
    Benchmark
    & \multicolumn{1}{c}{Queries}
    & \multicolumn{1}{c}{DPSize}
    & \multicolumn{1}{c}{GEQO}
    & \multicolumn{1}{c}{MCTS-E} \\
    \midrule
    JOB         & 113 &  61 &  52 & \textbf{ 79} \\
    JOB-Complex &  29 &  14 &  17 & \textbf{ 23} \\
    IMDb-CEB    & 200 & 121 & 115 & \textbf{154} \\
    \midrule
    Total       & 342 & 196 & 184 & \textbf{256} \\
    \bottomrule
  \end{tabular*}
\end{table}

\subsection{JOB}

\begin{figure}[!t]
  \centering
  \includegraphics[width=\linewidth]{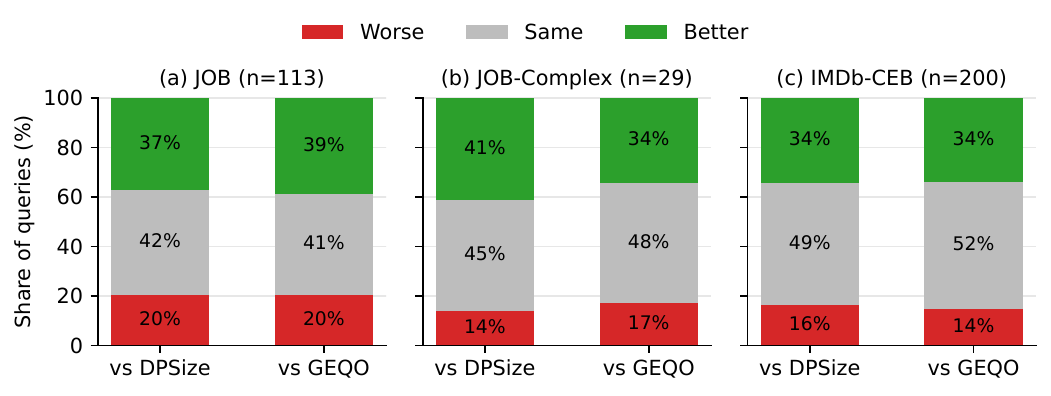}
  \caption{Per-query outcome distribution.}
  \label{fig:outcomes}
\end{figure}

Figure~\ref{fig:outcomes}(a) classifies each of the 113 JOB
queries into \emph{better}, \emph{same}, or \emph{worse} based
on its mean e2e time relative to DPSize and to GEQO, with a
$\pm 5\%$ tie band.  Most small JOB queries are effectively solved
by PostgreSQL's exact search, so the optimizers often agree.  The
separation appears after the GEQO threshold: on the high-arity JOB
slice, MCTS-Extreme avoids several poor GEQO plans while retaining
similar planning overhead.

\begin{figure}[!t]
  \centering
  \includegraphics[width=\linewidth]{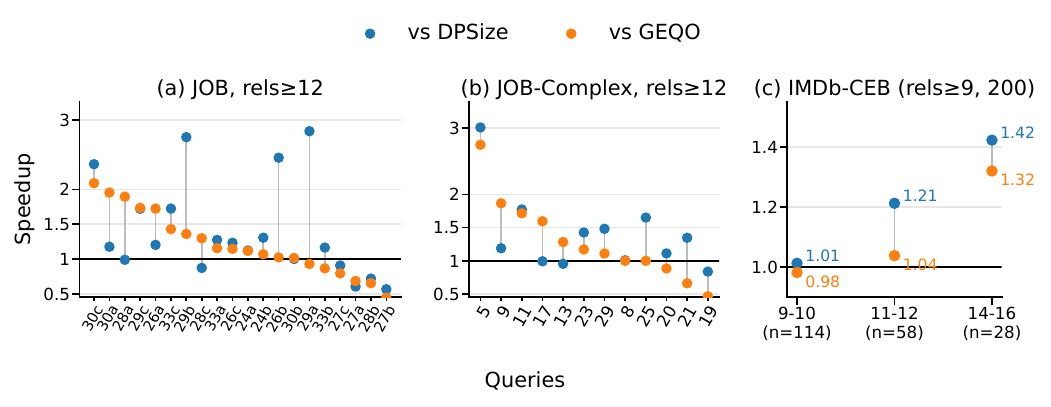}
  \caption{High-arity and arity-bucket speedups.}
  \label{fig:results_detail}
\end{figure}

\subsection{JOB-Complex (Stress Test)}
\label{sec:jobcomplex}

JOB-Complex contains 30 queries on the same IMDb schema as JOB,
with harder join structures.  We use it as a related stress test:
the database is shared with JOB, but the query templates differ.
The learned-baseline failures on this workload are discussed in
Section~\ref{sec:motivation}; they are not a controlled
distribution-shift experiment.

Figure~\ref{fig:outcomes}(b) shows the outcome
classification on JOB-Complex.  We exclude query~12, a
$\geq 12$-relation query on which MCTS-Extreme and GEQO exceeds the
5-minute \texttt{statement\_timeout}, so the
analysis below covers the remaining 29 queries.

This workload represents the primary cautionary case. MCTS-Extreme improves several high-arity queries and is clearly competitive with DPSize. The comparison with GEQO is less conclusive, however, given the limited number of high-arity queries in this workload. Query~12 further demonstrates that stochastic search methods relying solely on the optimizer cost model may fail to find a good plan within a reasonable time budget. In this case, cost degradations of less than 1\% can lead to orders-of-magnitude differences in execution time, revealing a substantial mismatch between estimated and actual performance. We therefore treat JOB-Complex primarily as evidence of robustness limitations rather than as an independent win over GEQO.

\subsection{Comparison to Learned MCTS Optimizers}
\label{sec:learned-comparison}

We next report results for the two learned MCTS optimizers we were
able to study, AlphaJoin and HyperQO.  Because neither system runs
as an in-engine optimizer, we use the reproducibility setup from
Section~\ref{sec:motivation}: each learned optimizer emits a join
order, which is executed through PostgreSQL and timed end-to-end
against the same default-PostgreSQL baseline used throughout this
section.
Table~\ref{tab:learned_outcomes} reports the per-query outcome
distribution under the same $\pm 5\%$ e2e tie band used in
Figure~\ref{fig:outcomes}.

Under this setup, the learned baselines fail for different reasons.
AlphaJoin's re-trained model produces systematically bad JOB plans
and cannot be applied to JOB-Complex because the encoded feature
dimensionality changes.  HyperQO is safer in the sense that it often
falls back to PostgreSQL's default plan, but the external steering
path still adds latency and produces no usable JOB-Complex hints.
These are deployability failures, not evidence that the original
methods are intrinsically worse search algorithms.  The relevant
contrast is architectural: MCTS-Extreme remains inside PostgreSQL's
planner and therefore avoids the hint-rewriting, model-drift, and
out-of-process inference boundary that dominate these reproductions.

\begin{table}[!t]
  \centering
  \small
  \setlength{\tabcolsep}{0pt}
  \caption{Learned-optimizer outcomes against PostgreSQL.}
  \label{tab:learned_outcomes}
  \begin{tabular*}{\columnwidth}{@{\extracolsep{\fill}}lrrrrr@{}}
    \toprule
    & \multicolumn{3}{c}{Per-query outcome} &
    \multicolumn{2}{c}{Total e2e impr.} \\
    \cmidrule(lr){2-4} \cmidrule(lr){5-6}
    & \multicolumn{1}{c}{Worse}
    & \multicolumn{1}{c}{Same}
    & \multicolumn{1}{c}{Better}
    & \multicolumn{1}{c}{All}
    & \multicolumn{1}{c}{$\geq 12$ rel} \\
    \midrule
    \multicolumn{6}{l}{\emph{JOB} (113 queries)} \\
    AlphaJoin    & $100.0\%$ & $0.0\%$  & $0.0\%$  & $-2366\%$ & $-976\%$  \\
    HyperQO      & $87.6\%$  & $11.5\%$ & $0.9\%$  & $-16.0\%$ & $-40.5\%$ \\
    MCTS-Extreme & $21.2\%$  & $42.5\%$ & $36.3\%$ & $+12.6\%$ & $+32.0\%$ \\
    \midrule
    \multicolumn{6}{l}{\emph{JOB-Complex} (29 queries)} \\
    AlphaJoin    & \multicolumn{5}{c}{did not run (feature-dim.\ mismatch)} \\
    HyperQO      & $58.6\%$  & $41.4\%$ & $0.0\%$  & $-4.3\%$  & $-3.3\%$  \\
    MCTS-Extreme & $20.7\%$  & $41.4\%$ & $37.9\%$ & $+4.4\%$  & $+12.7\%$ \\
    \bottomrule
  \end{tabular*}
\end{table}

\subsection{IMDb-CEB (High-Arity Workload)}
\label{sec:ceb}

IMDb-CEB \cite{negi2023robust} is a $3{,}000$-query workload
over the same IMDb schema as JOB, generated from CEB
templates so that every query joins many base relations with
deliberately correlated predicates.  We use a $200$-query subset
drawn randomly from the published corpus
with seed $42$ and restricted to queries with $\geq 9$ joined
relations (script and manifest included with the released
artifact).  This subset straddles PostgreSQL's GEQO threshold:
some queries remain in the exact-search regime, while the largest
queries exercise the randomized planner.

Figure~\ref{fig:results_detail}(c) shows the main trend.  The
advantage is not uniform across IMDb-CEB; it grows with relation
count.  Near the threshold, PostgreSQL's exact planner and GEQO
already find plans close to MCTS-Extreme's.  On the largest CEB
queries, the wider stochastic search pays off more often.  At the
same time, Figure~\ref{fig:outcomes}(c) shows a nontrivial tail of
queries where MCTS-Extreme loses, so the result is an aggregate
shift rather than a per-query dominance claim.

The clearest failure mode is template-specific.  CEB template
\texttt{11b} contains cases where the PostgreSQL cost model rates
MCTS-Extreme's chosen plans as essentially equivalent to DPSize's,
yet the plans execute slower.  This is the same cost/runtime
disconnect that makes search useful on other templates, but here it
cuts against MCTS-Extreme.

\subsection{Planning Time Within Response Time}

End-to-end response time includes planning overhead, so a better
join order is useful only if the search cost does not erase the
execution gain.  Tables~\ref{tab:workload_aggregates} and
\ref{tab:high_arity_aggregates} show that MCTS-Extreme stays in the
same planning-time range as GEQO on high-arity slices and remains
well below exact DPSize.  Planning overhead is therefore not the
source of the aggregate speedups, nor does it overturn them.  It can
matter on very short-running individual queries, which motivates
future per-query routing between optimizer modes.

\subsection{Configuration Ablation}
\label{sec:ablation}

Table~\ref{tab:ablation} varies MCTS-Extreme's own knobs only; no
baseline appears.  It covers both the full $200$-query workload and
the high-arity relations${\geq}12$ slice.  A seed-run is one
query-seed execution.  Each cell is a workload total over completed
seed-runs; the \textsc{to} column counts seed-runs that timed out
and are therefore excluded from the totals.  The table varies four
knobs: reward map, selection rule, plan shape, and search budget.
The reward-map rows were
recorded under an older shallow budget, so they diagnose failure
modes but are not a clean same-budget reward-map comparison.

\begin{table*}[!t]
  \centering
  \small
  \setlength{\tabcolsep}{0pt}
  \caption{IMDb-CEB configuration ablation.}
  \label{tab:ablation}
  \begin{tabular*}{\textwidth}{@{\extracolsep{\fill}}llrrrrrrrrr@{}}
    \toprule
    Reward $\phi$ & Selection
    & \multicolumn{1}{c}{$K$}
    & \multicolumn{1}{c}{$d$}
    & \multicolumn{1}{c}{$k$}
    & \multicolumn{1}{c}{\textsc{cost}}
    & \multicolumn{1}{c}{\textsc{plan}}
    & \multicolumn{1}{c}{\textsc{runtime}}
    & \multicolumn{1}{c}{\textsc{e2e}}
    & \multicolumn{1}{c}{\textsc{iter}}
    & \multicolumn{1}{c}{\textsc{to}} \\
    & & & & &
    \multicolumn{1}{c}{($10^6$)}
    & \multicolumn{1}{c}{(s)}
    & \multicolumn{1}{c}{(s)}
    & \multicolumn{1}{c}{(s)}
    & \multicolumn{1}{c}{($10^3$)}
    & \\
    \midrule
    \multicolumn{11}{l}{\emph{Full workload} (200 queries)} \\
    $-\log\widehat{C}$        & UCT-Extreme  & 1 & 10 &  9 & 45.6 &
    33.7 &  7014 &  7048 & 85.4 &  0 \\
    $\boldsymbol{-\log\widehat{C}}$ & \textbf{UCT-Extreme} &
    \textbf{1} & \textbf{10} & \textbf{10}
    & \textbf{45.6} & \textbf{33.9} & \textbf{7103} & \textbf{7137} &
    \textbf{85.7} & \textbf{0} \\
    $-\log\widehat{C}$        & UCT-Extreme  & 1 & 10 & 12 & 45.6 &
    33.9 &  7132 &  7166 & 86.2 &  0 \\
    $-\log\widehat{C}$        & UCT (mean)  & 1 & 10 & 10 & 46.2 &
    35.3 &  7140 &  7175 & 83.9 &  0 \\
    $-\log\widehat{C}$        & UCT-Extreme  & 1 & 10 & 11 & 45.7 &
    34.0 &  7197 &  7231 & 85.9 &  0 \\
    $-\log\widehat{C}$        & UCT-Extreme  & 1 & 10 &  8 & 45.6 &
    33.7 &  7200 &  7234 & 86.3 &  0 \\
    norm.\ $-\log\widehat{C}$ & UCT-Extreme  & 1 &  4 &  5 & 46.0 &
    29.5 &  7249 &  7279 & 40.6 &  0 \\
    $-\log\widehat{C}$        & UCT-Extreme  & 1 &  4 &  5 & 46.1 &
    28.0 &  7328 &  7356 & 30.6 &  0 \\
    $-\widehat{C}$            & UCT-Extreme  & 1 &  4 &  5 & 48.2 &
    24.8 &  7575 &  7600 &  3.9 &  0 \\
    $-\log\widehat{C}$        & UCT-Extreme  & 2 & 10 & 10 & 71.0 &
    33.8 &  9613 &  9647 & 79.4 &  4 \\
    $-\log\widehat{C}$        & UCT-Extreme  & 3 & 10 & 10 &111.0 &
    32.3 & 10462 & 10494 & 82.2 & 26 \\
    \midrule
    \multicolumn{11}{l}{\emph{Queries with $\geq 12$ relations} (57 queries)} \\
    $-\log\widehat{C}$        & UCT-Extreme  & 1 & 10 &  9 & 11.7 &
    12.2 & 2476 & 2488 & 27.4 &  0 \\
    $\boldsymbol{-\log\widehat{C}}$ & \textbf{UCT-Extreme} &
    \textbf{1} & \textbf{10} & \textbf{10}
    & \textbf{11.7} & \textbf{12.2} & \textbf{2562} & \textbf{2575} &
    \textbf{27.4} & \textbf{0} \\
    norm.\ $-\log\widehat{C}$ & UCT-Extreme  & 1 &  4 &  5 & 11.9 &
    10.4 & 2591 & 2602 & 15.7 &  0 \\
    $-\log\widehat{C}$        & UCT-Extreme  & 1 & 10 & 12 & 11.7 &
    12.2 & 2595 & 2607 & 27.4 &  0 \\
    $-\log\widehat{C}$        & UCT (mean)  & 1 & 10 & 10 & 12.4 &
    12.6 & 2629 & 2642 & 27.4 &  0 \\
    $-\log\widehat{C}$        & UCT-Extreme  & 1 & 10 & 11 & 11.7 &
    12.3 & 2647 & 2659 & 27.4 &  0 \\
    $-\log\widehat{C}$        & UCT-Extreme  & 1 & 10 &  8 & 11.7 &
    12.1 & 2669 & 2681 & 27.4 &  0 \\
    $-\log\widehat{C}$        & UCT-Extreme  & 1 &  4 &  5 & 11.9 &
    9.9 & 2679 & 2689 & 13.7 &  0 \\
    $-\widehat{C}$            & UCT-Extreme  & 1 &  4 &  5 & 12.1 &
    8.1 & 2709 & 2717 &  1.3 &  0 \\
    $-\log\widehat{C}$        & UCT-Extreme  & 2 & 10 & 10 & 15.1 &
    12.4 & 3419 & 3432 & 26.9 &  4 \\
    $-\log\widehat{C}$        & UCT-Extreme  & 3 & 10 & 10 & 20.2 &
    11.2 & 3488 & 3499 & 25.2 & 19 \\
    \bottomrule
  \end{tabular*}
\end{table*}

\begin{table}[!t]
  \centering
  \small
  \setlength{\tabcolsep}{0pt}
  \caption{Linear/zig-zag search controls on IMDb-CEB queries with
  at least 12 joined relations.}
  \label{tab:search_control}
  \begin{tabular*}{\columnwidth}{@{\extracolsep{\fill}}lrrrrrr@{}}
    \toprule
    Search
    & \multicolumn{1}{c}{\textsc{cost}}
    & \multicolumn{1}{c}{\textsc{plan}}
    & \multicolumn{1}{c}{\textsc{runtime}}
    & \multicolumn{1}{c}{\textsc{e2e}}
    & \multicolumn{1}{c}{\textsc{iter}}
    & \multicolumn{1}{c}{\textsc{to}} \\
    & \multicolumn{1}{c}{($10^6$)}
    & \multicolumn{1}{c}{(s)}
    & \multicolumn{1}{c}{(s)}
    & \multicolumn{1}{c}{(s)}
    & \multicolumn{1}{c}{($10^3$)}
    & \\
    \midrule
    MCTS-E             &  7.6 & 3.2 & 429.3 & 432.6 & 6.7 &  0 \\
    SAIO   & 28.0 & 0.9 & 795.9 & 796.9 & 6.7 & 37 \\
    II & 57.0 & 0.4 & 941.6 & 942.0 & 6.7 & 61 \\
    \bottomrule
  \end{tabular*}
\end{table}

The ablation points to plan shape as the load-bearing choice.
Allowing more bushy structure increases search cost and introduces
timeouts on this workload, while the $K{=}1$ linear/zig-zag mode is
stable.  Within that space, the remaining knobs move the result much
less.  UCT-Extreme lowers the search objective, but its end-to-end
advantage over classical mean-UCT is not statistically established at the
evaluated budget.  The top-$k$ sweep is similarly flat near the
chosen default.  We therefore treat UCT-Extreme as a useful search
statistic, not as the sole driver of the measured speedups.

Table~\ref{tab:search_control} isolates the search algorithm from
the $K{=}1$ linear/zig-zag plan-shape restriction.  It compares
MCTS against two non-tree local-search controls, such as Iterative Improvement (II) and Simulated Annealing (SAIO) at the same
$d{=}10$, $k{=}10$, and
$480$-iteration budget.  Because the local searches time out
frequently, finite timing totals use the common complete subset and
the timeout column reports failures on the full
relations${\geq}12$ slice.  The result is not just that MCTS is
faster on the filtered subset; it is that the tree search reaches the
full slice without the timeout behavior shown by the local searches.
This supports the
claim that tree-guided exploration adds value within the same
linear/zig-zag search space.

\section{Discussion}
\label{sec:discussion}

\paragraph{Where It Helps.}
MCTS-Extreme is not a replacement for exact dynamic programming on
small joins.  Its useful regime begins where PostgreSQL stops exact
enumeration and falls back to randomized GEQO search.  In that
regime, a modest MCTS budget can hedge against poor GEQO choices
without making planning the dominant cost.

\paragraph{Integration Lessons.}
The learned-baseline reproductions failed mainly at the database
boundary: hint rewriting, learned-model feature drift, and
out-of-process inference all mattered before the search algorithm
itself could be judged.  MCTS-Extreme makes a more conservative
tradeoff.  It does not replace PostgreSQL's cost model or executor;
it changes only how the planner searches join orders.

\paragraph{Training Cost.}
MCTS-Extreme has no training stage, so the relevant cost is the
per-query planning-plus-execution time reported in
Section~\ref{sec:experiments}.  Counting training for AlphaJoin or
HyperQO would make the deployability gap larger, but it is not
needed for the main comparison: the learned systems already fail to
provide a stable in-engine optimization path in our setup.

\paragraph{Limits.}
The workload evidence is still narrow.  JOB, JOB-Complex, and
IMDb-CEB are public real-data join-ordering benchmarks that reach
the high-arity regime, but all use the IMDb schema.  The empirical
claim should therefore be read as an IMDb-family result, not as
evidence of cross-schema generality.

The current extension also optimizes only the flat join list exposed
by PostgreSQL.  It reuses PostgreSQL's join methods, cost model, and
cardinality estimates, and it does not optimize subqueries that the
host planner leaves unflattened.  Per-query routing between exact
DPSize, GEQO, and MCTS-Extreme remains future work.

\section{Reproducing Our Results}

The MCTS-Extreme source tree, the parameter settings used for each
experiment, the JOB and JOB-Complex query corpora, and the analysis
notebook for the plots are released through an anonymized git
repository at
\url{https://github.com/ooaarg/tfmctsqo}.

\section{Conclusion}

MCTS-Extreme explores a conservative point in the join-optimizer
design space: keep PostgreSQL's cost model and executor, but replace
the high-arity join-order search with a bounded MCTS procedure.  The
experiments suggest that this tradeoff is useful when PostgreSQL
leaves exact DPSize search and relies on GEQO.  The main driver is
the stable $K{=}1$ linear/zig-zag search space plus tree-guided
exploration; the UCT-Extreme statistic improves the search objective
but does not by itself explain the measured end-to-end gains.  The
result is a deployable, training-free optimizer mode with clear
limits: it helps in the high-arity regime, and a
per-query router should send queries to exact planning or a different
plan shape where those are better.



\bibliographystyle{ACM-Reference-Format}
\bibliography{refsv2}


\end{document}